\documentclass[showpacs,amssymb,aps]{revtex4}

\newcommand{\be}{\begin{equation}}
\newcommand{\ee}{\end{equation}}

\usepackage{graphicx}

\begin{document}

\title{On the energy-momentum tensor in non-commutative gauge theories}  

\author{Ashok Das$^a$ and  J. Frenkel$^b$}
\affiliation{$^a$Department of Physics and Astronomy,
University of Rochester,
Rochester, NY 14627-0171, USA}
\affiliation{$^b$ Instituto de F\'{\i}sica,
Universidade de S\~ao Paulo,
S\~ao Paulo, SP 05315-970, BRAZIL}

\bigskip

\begin{abstract}

We study the properties of the energy-momentum tensor in
non-commutative gauge theories by coupling them to a
weak external gravitational field. In particular, we show that the
stress tensor of such a theory coincides exactly with that derived
from a theory where
a Seiberg-Witten map has been implemented (namely, the procedure is
commutative). Various other interesting features are also discussed.
\end{abstract}


\maketitle

Non-commutative field theories \cite{seiberg,douglas} have been of
much 
interest in  recent years following developments in superstring
theory. In such theories, the natural  parameter of non-commutativity,
$\theta^{\mu\nu}$, is a dimensionful quantity. It is assumed to be
a  constant anti-symmetric tensor and this leads to a violation of
Lorentz invariance in such theories, which also puts a bound on
the  magnitude of the parameter of non-commutativity
\cite{carrol}. In view  of  these,
it is expected that scale invariance is also violated and
correspondingly, the  energy-momentum tensor has been studied
in non-commutative gauge theories from several points of view
\cite{bichl,grimstrup,kruglov,dorn}. 

In non-commutative gauge theories, there is a map known as the Seiberg-Witten
map \cite{seiberg}, which ensures the gauge equivalence between the
non-commutative theory and that of an ordinary gauge theory defined on
a commutative space. A surprising outcome of these
studies \cite{grimstrup}, which use an improved Belinfante method
\cite{jackiw}, is that the energy momentum tensor of the
non-commutative theory, upon using the Seiberg-Witten map, does not
coincide with the energy-momentum tensor of the theory obtained from
this map. Namely, the Seiberg-Witten map and the derivation of the
energy-momentum tensor do not seem to commute. In this brief report,
we study this question by coupling the non-commutative
gauge 
theory (with and without the Seiberg-Witten map) to gravity
\cite{das,ooguri,barcelos}. We show that the energy-momentum tensors 
derived by this method do
coincide, namely the map commutes with the derivation of the
energy-momentum tensor. The energy-momentum tensors are shown to be
manifestly symmetric and traceless, but conserved covariantly. It is
shown that it is possible to obtain, from this, a conserved tensor
which, however, is neither symmetric and traceless and coincides with
the one obtained from the improved Belinfante method. We shall discuss
here, for simplicity, only the
non-commutative $U(1)$ gauge theory, although the above conclusions
naturally extend to the $U(N)$ case.

The non-commutative $U(1)$ theory is described by the gauge invariant
action
\begin{equation}
S = - \frac{1}{4}\,\int d^{4}x\,\hat{F}_{\mu\nu}\star
\hat{F}^{\mu\nu}\label{1}
\end{equation}
where the star product of two functions is defined by
\begin{equation}
A(x)\star B(x) =
e^{\frac{i}{2}\theta^{\mu\nu}\partial_{\mu}^{\eta}\partial_{\nu}^{\xi}}
  \left.A(x+\eta) B(x+\xi)\right|_{\eta=\xi=0}\label{2}
\end{equation}
and the field strength tensor has the form
\begin{equation}
\hat{F}_{\mu\nu} = \partial_{\mu}\hat{A}_{\nu} -
\partial_{\nu}\hat{A}_{\mu} -i(\hat{A}_{\mu}\star \hat{A}_{\nu} -
\hat{A}_{\nu}\star \hat{A}_{\mu})\label{3}
\end{equation}
where $\hat{A}_{\mu}$ denotes the $U(1)$ gauge field defined on a
non-commutative space. The Euler-Lagrange equations following from
(\ref{1})  lead to
\begin{equation}
\hat{D}_{\mu}\star \hat{F}^{\mu\nu} = \partial_{\mu}\hat{F}^{\mu\nu} -
i (\hat{A}_{\mu}\star \hat{F}^{\mu\nu} - \hat{F}^{\mu\nu}\star
\hat{A}_{\mu}) = 0\label{4}
\end{equation}

By coupling the theory in (\ref{1}) to a weak external gravitational
field, we obtain
\begin{equation}
S = - \frac{1}{4}\,\int d^{4}x\,\sqrt{-g}\,\star g^{\mu\lambda}\star
g^{\nu\rho}\star \hat{F}_{\mu\nu}\star \hat{F}_{\lambda\rho}\label{5}
\end{equation}
where $g$ denotes the determinant of the metric with a signature
$(+,-,-,-)$. The energy-momentum tensor of the theory can now be obtained as
\begin{equation}
\hat{T}_{\mu\nu} = \left.\frac{2}{\sqrt{-g}}\,\frac{\delta S}{\delta
  g^{\mu\nu}}\right|_{g^{\mu\nu}=\eta^{\mu\nu}}\label{6}
\end{equation}
We note that the star product involving the metric is
not of consequence since the metric is set to the Minkowski metric at
the end. It follows now in a straight forward manner that, for the
theory in (\ref{1}),
\begin{equation}
\hat{T}_{\mu\nu} = -\frac{1}{2} \left(\hat{F}_{\mu\lambda}\star
\hat{F}_{\nu}^{\,\lambda} + \hat{F}_{\nu\lambda}\star
\hat{F}_{\mu}^{\,\lambda} - \frac{1}{2}\,
\eta_{\mu\nu}\,\hat{F}_{\lambda\rho}\star
\hat{F}^{\lambda\rho}\right)\label{7}
\end{equation}
This tensor is manifestly symmetric and traceless and coincides with
the one obtained through the conventional derivation in
\cite{grimstrup}. Furthermore,  using the
equations of motion (\ref{4}) as well as Bianchi identity, it can be
checked that this tensor is covariantly conserved, namely,
\begin{equation}
\hat{D}_{\mu}\star \hat{T}^{\mu\nu} = 0\label{8}
\end{equation}

In this theory, the Seiberg-Witten map, to leading order in
$\theta^{\mu\nu}$,  leads to the correspondence 
\begin{eqnarray}
\hat{A}_{\mu} & = & A_{\mu} - \frac{1}{2}\,\theta^{\alpha\beta}
A_{\alpha} (\partial_{\beta} A_{\mu} + F_{\beta\mu}) +
O(\theta^{2})\nonumber\\
\noalign{\vskip 4pt}%
\hat{F}_{\mu\nu} & = & F_{\mu\nu} - \theta^{\alpha\beta}
(F_{\mu\alpha}F_{\beta\nu} + A_{\alpha}\partial_{\beta} F_{\mu\nu}) +
O(\theta^{2})  =
F_{\mu\nu} - \left((F\theta F)_{\mu\nu} + (A\theta\partial)
F_{\mu\nu}\right) + O(\theta^{2})\label{9}
\end{eqnarray}
where we have used an obvious matrix notation for
simplicity. Here, the variables without a ``hat'' correspond to
quantities defined on a commutative space and correspondingly
\begin{equation}
F_{\mu\nu} = \partial_{\mu}A_{\nu} - \partial_{\nu}A_{\mu}\label{10}
\end{equation}
Substituting the map (\ref{9}) into (\ref{1}), we obtain, to order
$\theta$, 
\begin{equation}
S = - \frac{1}{4}\,\int d^{4}x\,\left[(1 -
  \frac{1}{2}\,\theta^{\alpha\beta}F_{\alpha\beta})\,F_{\mu\nu}F^{\mu\nu}
  + 2 {\rm Tr}\,(\theta F^{3}) - \partial_{\beta}
  \left(\theta^{\alpha\beta} A_{\alpha}
  F_{\mu\nu}F^{\mu\nu}\right)\right]\label{11}
\end{equation}
Although a total divergence does not contribute to the equations of
motion and  is normally neglected in flat
space-time, it does contribute to the energy-momentum tensor
\cite{das} and that is the reason why we have manifestly kept such a
term in (\ref{11}). The Euler-Lagrange equations following from this
action lead to 
\begin{equation}
\partial_{\mu}\left[(1-\frac{1}{2}\theta^{\alpha\beta}F_{\alpha\beta})\,
F^{\mu\nu} - \left((F^{2}\theta)^{\mu\nu} + (F\theta F)^{\mu\nu} +
(\theta F^{2})^{\mu\nu}\right) -
\frac{1}{4}\,\theta^{\mu\nu}\,F_{\lambda\rho}F^{\lambda\rho}\right] =
0\label{12}
\end{equation}
Furthermore, by coupling (\ref{11}) to a weak external gravitational
field, we can derive the energy-momentum tensor, as in (\ref{6}), to
obtain
\begin{eqnarray}
T_{\mu\nu} & = & -
(1-\frac{1}{2}\theta^{\alpha\beta}F_{\alpha\beta})\,
\left(F_{\mu\lambda}F_{\nu}^{\,\lambda}
-\frac{1}{4}\,\eta_{\mu\nu}\,F_{\lambda\rho}F^{\lambda\rho}\right) -
\left((F\theta F^{2})_{\mu\nu} + (F\theta F^{2})_{\nu\mu} -
\frac{1}{2}\,\eta_{\mu\nu}\,{\rm Tr} (F\theta F^{2})\right)\nonumber\\
\noalign{\vskip 4pt}%
 &  & \quad + \partial_{\beta} \left(\theta^{\alpha\beta}
A_{\alpha} (F_{\mu\lambda}F_{\nu}^{\,\lambda} -
\frac{1}{4}\,\eta_{\mu\nu}\,F_{\lambda\rho}
F^{\lambda\rho})\right)\label{13}
\end{eqnarray}

We note that the energy-momentum tensor is manifestly symmetric and
traceless as is $\hat{T}_{\mu\nu}$ in (\ref{7}). Furthermore, it can
be easily checked that, under the Seiberg-Witten map in (\ref{9}) to
leading order in $\theta$,
\begin{equation}
\hat{T}_{\mu\nu} = T_{\mu\nu}\label{14}
\end{equation}
Namely, the energy-momentum tensor of the non-commutative theory,
under a Seiberg-Witten map, goes over to that obtained from a theory
with the map so that the process is commutative. Furthermore, under a
Seiberg-Witten map to the leading order in $\theta$, eq. (\ref{4}) gives
\begin{equation}
\partial_{\mu}\left[
  (1-\frac{1}{2}\theta^{\alpha\beta}F_{\alpha\beta}) F^{\mu\nu} -
  (F\theta F)^{\mu\nu} - (\theta F^{2})^{\mu\nu}\right] = 0\label{15}
\end{equation}
The dynamical equations (\ref{12}) and (\ref{15}) do not seem to
coincide at first sight. However, it is easy to check, with the help
of the relation (which holds to linear order in $\theta$),
\begin{equation}
\partial_{\mu}\left[(F^{2}\theta)^{\mu\nu} +
  \frac{1}{4}\,\theta^{\mu\nu}\,F_{\lambda\rho}F^{\lambda\rho}\right]
  = 0\label{16}
\end{equation}
that the dynamical equations following from the non-commutative
theory go over, under a Seiberg-Witten map, to the ones following from
the  deformed theory under the same map.

Since the dynamical equations as well as the energy-momentum tensors
of the two theories are the same (to linear order in $\theta$), it
follows from (\ref{8}) that
\begin{equation}
\partial_{\mu} T^{\mu\nu} = \partial_{\mu}\left[(\theta FT^{(0)})^{\mu\nu} -
  \partial_{\beta} (\theta^{\alpha\beta} A_{\alpha}
  T^{(0)\,\mu\nu})\right]\label{17}
\end{equation}
where $T^{(0)\,\mu\nu}$ corresponds to the energy-momentum tensor with
$\theta=0$. Thus, we see that the energy-momentum tensor of the
Seiberg-Witten  deformed theory
is not conserved in the ordinary sense, which is a reflection of the
covariant conservation of the stress tensor in the non-commutative
theory. However, it is clear that we can define a modified energy-momentum
tensor
\begin{eqnarray}
\overline{T}^{\mu\nu} & = & T^{\mu\nu} - (\theta FT^{(0)})^{\mu\nu} +
\partial_{\beta} (\theta^{\alpha\beta} A_{\alpha}
T^{(0)\,\mu\nu})\nonumber\\
\noalign{\vskip 4pt}%
 & = & - (1-\frac{1}{2}\theta^{\alpha\beta} F_{\alpha\beta})
(F^{\mu\lambda}F^{\nu}_{\,\lambda} -
\frac{1}{4}\,\eta^{\mu\nu}\,F_{\lambda\rho}F^{\lambda\rho}) -
\left((F\theta F^{2})^{\mu\nu} + (F\theta F^{2})^{\nu\mu} -
\frac{1}{2} \eta^{\mu\nu} {\rm Tr}\,(F\theta F^{2})\right)\nonumber\\
\noalign{\vskip 4pt}%
 &  & \qquad - (\theta F T^{(0)})^{\mu\nu}\label{18}
\end{eqnarray}
which will be conserved, namely,
\begin{equation}
\partial_{\mu} \overline{T}^{\mu\nu} = 0\label{19}
\end{equation}
However, this tensor is no longer symmetric or traceless because of
the last term in (\ref{18}). Furthermore, it 
coincides with the energy-momentum tensor derived from the improved
Belinfante method.

\vskip 1cm

\noindent{\bf Acknowledgment:}

This work was supported in part by US DOE Grant number DE-FG
02-91ER40685, by CNPq and FAPESP, Brasil.

\end{document}